\documentclass[12pt]{article}

\usepackage{amsmath}    
\usepackage{graphicx}   
\usepackage{verbatim}   
\usepackage{color}      
\usepackage{hyperref}   
\usepackage{braket}

\begin{document}

\title{\bf Causal Event Networks: Cognition, Complexity and Physical Laws}

\author{Vahid R. Ramezani}

\date { Feb. 10, 2014}

\maketitle

\begin{abstract}

Information flow framed in a computational and complexity  context is relevant to the understanding of cognitive
processes and awareness. In this paper, we begin with analyzing an information theory   framework 
developed in recent years under Information and Integration Theory (IIT) based on interactions among partitions of cognitive information sets. We discuss the scope and limitations of these ideas, introducing a related measure for partitioning information sets. We introduce a set of postulates describing cognition as a partially ordered set of events in space and time. We consider the relevant sequential and concurrent computational concepts in an idealized minimal cognitive device. The concept of fundamental cognitive chain formalizes temporal limits of cognition.

\end{abstract}

\section*{Introduction}

Cognitive processes, and hence our subjective experience can be viewed as transformation of information in space and time. Exploring the relationship between matter-energy and spacetime led us from the development of Classical Mechanics and Electro-Magnetism  to the Standard Model of Physics and General Relativity. Holographic Principle \cite{Hooft}  and other ideas describing the universe in terms of information and causality, further, confirm the longstanding philosophical perspective that the information  content of the universe is unified.  

Because human brain is an information processing system, the application of statistical information theory associated with neural activities  might be an effective approach. Another approach is the direct application of the coherent structure of Quantum Mechanics (QM). The QM approach is attractive because it  addresses the hard problem of consciousness: objects and their qualities constituting the content of our subjective experience appear unified in a cognitive background \cite{Penrose1}-\cite{Penrose2}. However, Max Tegmark has shown that the brain's thermal properties probably prevent  a meaningful QM wave function to have a significant role in the information processes of the brain \cite{Tegmark}. 
In this paper, we propose a framework based on causal temporal network of events. We propose a number of postulates and follow their consequences.

According to IIT \cite{Tononi}, the binding problem and consciousness itself can be understood by a statistical simultaneous integration and differentiation measure. In our view, the concept of  causality of events is fundamental together with development of a functional geometry where cognitive computation is the central element. We view cognition as a transformation and hence measures of complexity, in our view, should be directly defined on computational representations, for example on complex directed graphs. 

Causality is an important concept in physical systems. 
A remarkable property of Minkowski space is that its geometry can be recovered from its causal order \cite{Robb}.
The situation in curved spacetime is more complex. But the geometry is still recoverable with additional assumptions. For example, it is shown  in \cite{Martin} not only the topology, but the geometry of hyperbolic spacetime can be reconstructed order theoretically from its causal structure together with an appropriate measurement.  More recently, Smolin and Cores  have developed a preliminary model that shows spacetime itself can emerge from a casual structure  and that Quantum Mechanics can be partially obtained from 
a causal network \cite{Smolin1}-\cite{Smolin2}. 

Cognition, we postulate, occurs at classical scales and in fact at a level of abstraction corresponding to neural computations. A causal network arising from these computations holds the solution to the \lq\lq binding problem" \cite{Chalmers}. Classical computation does not conserve energy and thus cognitive causal structure is  different from networks in physics which typically  have explicit or implicit formulation of energy conservation. We cannot  model
consciousness as an open system in a straightforward way; open systems in physics can be understood as relations between one system and a larger system containing it. Consciousness  is not a component of a larger physical system.

We discuss information theocratic issues in  Section 1. The main contribution of this section is the introduction of a measure which formalizes of the concept of information contained in  subsets in the range of a transformation, \lq\lq parts", of a system about the \lq\lq whole", the domain of the transformation. We apply it to a local diffeomorphism in $R^n$ and show it reflects the complexity of manifolds generated by components.  We also  show the measure provides a solution to the \lq\lq weakest link" problem raised in  \cite {Tononi}. We give an example to demonstrate that coordinate dependent definition of IIT, leads to inconsistent results and describe possible solutions using our measure.  Causal Cognitive Networks are discussed in Section 2. A main objective of Section 2 is to describe possible constructive  mechanisms which would theoretically {\it require} consciousness to operate \lq\lq near present"   and thus cause its failure under time dilation (in the sense of slowing down all processes of the network at a uniform rate ). Placing a simple limit on the spread time cannot explain the mechanism by which this failure occurs. It cannot be explained by synchronization failure or anything that remains invariant under  uniform time dilation.   Our point of view is that no matter how complex  cognitive  processes, they must be temporally differentiated into separate computational clusters which operate sequentially; failure basically occurs because computational density  along causal paths  falls below a critical threshold. We also postulate that the number of  allowed sequential computations is limited. The twin limiting thresholds (in density and causal sequence)  limit the time spread. Roughly speaking, we propose a self-similar computational structure where more complex cognitive \lq\lq shapes" (whose causal paths determine the number of sequential computations) require no more time than the fundamental generating elements of the self-similarity. Furthermore, temporal segregation of computational clusters is defined constructively.

\section {Information Theory and Cognitive Integration}

If consciousness is the experience of transformation and integration of information, it is natural to study functions which integrate 
information from an information domain into an information range.

We begin with a simple but illuminating example as to what \lq\lq information integration" could mean.  Suppose we have a collection of subsets of points in $R^2$. Each subset $x$ is given by the points on the sides of a triangle  including the vertices and a fixed number of other points on each side  obtained by $s$-equipartition for some value $s > 1$. The triangles  might intersect. A triangle  can be uniquely identified by its three vertices. We can transmit to another party the identity of a particular triangle by transmitting the coordinates of the vertices from which all other points can be reconstructed by equipartition. We could transmit 6 points, 2 on each
side instead but that would not be as efficient. In fact, given $s$, 3 is the conditional Kolmogorov complexity 
\cite{Cover} of each triangle ( We could assume the vertices are given by pairs of integers in which case, to be precise, the complexity  would be the total number of bits required to describe a set of 6 integers, but we will not be precise with these details here). As a shorthand, let us say the Kolmogorov complexity is 3 corresponding to 3 components  or a triangle's degrees of freedom. How much information is contained in the components about the whole triangle? A single vertex gives us almost no information if $s$ is large. Any two vertices give us about a third of the total information by specifying one of the three sides. Knowing all coordinates of three vertices specifies the triangle. We will formulate this relation between the \lq\lq parts" and the \lq\lq whole" more generally later. But, first, let us explore the example further. Observe each vertex alone can specify  a single point from all possible points of the triangle. Summing the information about the vertices {\it individually}  would not  produce a significant increase; we will at most know three points. We have made an implicit assumption here  that $s$ and the method of integrating the information, the computational rules to produce the points of the equipartition, $f,$ from the vertices , are known to the receiving party; thus, the Kolmogorov complexity is conditional: 

$$k(x | f, s)=3=|\{v_1, v_2, v_3\}| $$

\noindent
We can make the \lq\lq integration" data explicit by considering  

$$k(x| f) = 4=|\{v_1, v_2, v_3, s\}|. $$ 

\noindent
How do various partitions of the information set $\{  v_1, v_2, v_3 , s\}$ inform us about the triangle?  

Partitions $\{  \{v_1, v_2, v_3 \}  \{ s\}    \}$ and $ \{ \{ v_1, v_2, s \} \{ v_3\}\} $ having matching cardinalities behave differently; the former produces almost no information
even though it contains all the three vertices, while the latter produces about a third of total information. This is because the second
partition combines structural data with  integration data. 

Consider the integration of visual information from two eyes that gives us our depth perception. Can we meaningfully find a partition
of the integrated information that maps to a natural bi-partition of the  original information pattern? Depth perception is
generated from the slight information differences between the images that each eye produces, essentially, symmetric with respect to the two information sets produced by the eyes. 
Instead of triangles above consider circles identified by their center and radius $( c, r)$ . There is no natural partition of a circle
that matches  the natural bi-partition $( c, r)$ because both $c$ and $r$ are obtained from integration of information over the entire circle (from any three points). 

We next define IIT following the summary given in 
\cite{Barret}. This measure was defined for discrete, Markovian systems, i.e. systems with (i) a discrete set of possible states, and (ii) dynamics for which the current state depends only on the state at the previous time-step.
Let $X$ be a discrete, Markovian system. IIT measure, $\Phi_{DM}$, attempts to compare the information generated by the whole system to information generated by its parts, when the system transitions to a particular state $X_1=x$ from a proceeding state $x_0$ characterized with maximum entropy entropy distribution for the system. This is performed by use of KL divergence to compare (i) the conditional probability distribution for the preceding state of the whole given the current state; (ii) the joint distribution for the preceding states of parts given their respective current states. The {\it effective information}, $\phi_{DM}[X;x, \it P]$, generated by $X$ being in state $x$, with respect to Partition ${\it P=\{ M^1,...,M^N \}}$, is given by KL divergence:  

$$\phi_{DM}[X;x,P]=: D_{KL}\left( P_{X_0|X_1=x} \mid\mid \prod_{k=1}^{N}P_{M_0^k|M_1^k=m^k}\right)$$

\noindent
Here $m^k$ is the state of $k^{th}$ sub-systems of the partition when $X$ 
has state $x$.

The integrated information is defined as the effective information with respect to the minimum information partition (MIP). The MIP, $\it P^{MIP}(x)$,  is defined as the partition that minimizes the effective information when it is normalized by

$$ K_M \left ( \{ M^1,...,M^N \} \right )=: (N-1)min_k\left( [H(M_0^k)\right] ) $$

\noindent
Normalization is necessary because sub-systems that are almost as large as the whole system typically generate almost as much information as the whole system. Therefore, without normalization, most systems would have a highly imbalanced MIP, (e.g., one element versus the remainder of the system) and a trivially small value for integrated information. The normalization  ensures that integrated information is specified using a partition defined using a weighted minimization of the effective information, with a bias towards partitions into sub-systems of roughly equal size. 

Given the MIP, the integrated information $\phi_{DM}(X;x)$ generated by the system $X$ entering state $x$ is simply the non-normalized effective information with respect to the MIP,

$$\Phi_{DM}[X;x]=\Phi_{DM}[X;x, \it P^{MIP}(x)]$$

\noindent
 IIT assumes there is an optimal coordinate matching, then (for  the optimal partition) places the information from the elements of the partition in the range back into the corresponding elements of partition in the domain by probabilistic conditioning. It then compares the information about the whole of the domain which would have been generated if the partitions in the domain were allowed to \lq\lq communicate" vs. information which is generated if the partitions where to be used separately in describing the whole. 
 
How is the whole divided into parts? More precisely if we partition an information set after a transformation, how much information does a single element of the partition in the range contain about the entire domain? We can also compare the average information contained in (the elements of) one partition in the range about the entire domain.

Note that in the previous example, we focused on how much of a triangle we can identify, not how well we can discriminate one triangle from
the other. The two problems are related but they are not the same. By reconstructing a certain part of
an object we can increase the ability to recognize it.

We give an example to clarify the situation. Consider a local diffemorphism,  $ f: R^n \rightarrow R^n, $  $ f(x_0)= 0$ defined on the open set $x_0 \in U$. Each component  $ f_i: R^n \rightarrow R$ is a submersion. It is clear that each component when set to zero, defines a manifold of dimension $n-1$. Each manifold (the part)  contains information about the solution (the whole). Generally, the intersection of two manifolds defines a manifold of lower order. The intersection of all manifolds has dimension zero, which is the  solution $x_0$. We can express this in probabilistic language of IIT by 
filling $U$ with random points defining a random grid. Initially, the solution could be any of the points in the volume and hence a uniform probability is generated, each manifold excludes all the points which do not intersect with it and hence reduces the space of possible solutions to a uniform distribution over grid points on the manifold only. Given a coordinate system, can we map the information in each component of the range to a particular component in the domain? Generally, this cannot be done meaningfully. For example, a linear manifold might \lq\lq cut across" the coordinate system and thus generate substantial information on several coordinate axes. If the function $f$ is affine, one might hope to define a preferred coordinate system, for example using some eigen space decomposition of the domain. However, for non-linear functions, the manifolds of lower order are non-linear (for example a section on the surface of a sphere). 

There are computational limits for an algorithmic implementations which must be considered. Generally, the number of points on each $i$-manifold depends on the volume of its intersection with $ U $. Given a limited degree of computational accuracy, an algorithm will generate a region of uncertainty around the manifold which depends on the machine accuracy and the particular set of calculations, denoted by $\epsilon (i) $.  Assume $ U $ is a hyper-cube with sides of length $ L $.  Therefor, the number of points $N$ associated with each manifold is  approximately the portion of points that fall into that volume (surface): $ \epsilon(i)V(i)N/L^n$. A  manifold of dimension $n-1$ inside this cube, has a volume (surface) proportional to $L^{n-1}$ given by a proportionality parameter $C(i)$ if it is sufficiently smooth. If we approximate $\epsilon(i)$ with $\epsilon$ uniformly, and fill up the volume of the cube of volume $L^n$ with approximately as many grid points
as computational accuracy allows $(L/\epsilon)^n$, the number of points associated with the manifold would be
 
 $$C(i)\epsilon^{-n+1}L^{n-1}=q$$

\noindent
Thus, knowing a component changes our knowledge of the solution from a uniform distribution $1/N$ to a distribution $ 1/q $ while knowledge of all components essentially produces a delta  probability density function at the solution.

How can we compare information contained in parts  of the range individually with the information that they provide together as a whole?
One way to do so is to look at the difference between the probability induced by a subset of the range vs. probability induce by the whole range about the entire domain. Assuming a discrete structure, let probability induced  by the information contained in the entire range be given by $p_i$. Consider a particular partition of the range into disjoint subsets and for each subset denote by $q_i^j$ the
probability induced on the elements of the domain by  $j^{th}$  partition. The information loss by restricting the available information can then be measured by KL-divergence:

$$D_{KL}(P|Q^j)= \sum_i p_i~ log \left({p_i\over q_i^j}\right )$$

The divergence is well defined because $q_i^j=0$ implies $p_i=0$;  we generally think of induced information in terms of set intersections or imposition of a consistent set of conditions. 
We can also compare the information contained in parts of a partition vs. information obtained from their union (the whole of the range) 
by averaging the KL-divergence over the elements of the partition:

\begin{align}
{\bar D} ( {\it P}) &= 1/N \sum_{j=1}^{N}\sum_i p_i ~ log \left ({p_i\over q_i^j}\right) \\
                       &= 1/N \sum_i p_i\sum_{j=1}^{N}  log \left ({p_i\over q_i^j}\right ) \\
                       &=  1/N \sum_i p_i~ log \left({p_i^N\over \prod_j^N q_i^j}\right) \\
                       &= \sum_i p_i~ log \left({p_i\over  {\bar q_i}}\right )
\end{align}                       
      
\noindent              
where $ {\bar q_i=(\prod_j^N q_i^j)^{1/N}}$ is the geometric average.  
For our manifold example, if we group the components into $N$
groups each containing $r_i$ components, after some calculations, we have
$${\bar D}= 1/N\sum_{i=1}^{N} log C(r_i)+ n(1-1/N)log(L/\epsilon)$$ 
Note for large $N=n$, $L=1$ the second term simplifies to $-nlog\epsilon$.  The second  term of the sum only depends on the size of the partition. The first term depends on the average deviation of  the volume (surface) from the cube of corresponding dimension. For a surface, It is a measure of complexity of  that surface with more \lq\lq valleys" and \lq\lq hills"  having larger surface relative to the square of the same dimension. For a cube one can imagine temporal isolations of a cube inside a spacetime hypercupe where each point of the cube is projected out to varying time intervals and so on for higher dimensions. 

\noindent
For random variables the same idea can be expressed in information theocratic terms as we show next.
\noindent
Given a set of random variables $D: X=\{x_1,... x_l\}$ defining our \lq\lq domain" and a set of random variables $R: Z=\{ z_1,... z_m\}$ defining the \lq\lq range", consider a partition  $P=\{ r_1,... r_N\}$ of the range and define the following

$$\psi(\it P)=\sum p(z_1,... z_m ) \sum p(x_1,... x_l|z_1,... z_m ) log {{ p( x_1,... x_l| z_1,... z_m)} \over { (\prod_j^N p( x_1,... x_l| r_j))^{1/N}}}$$

\noindent
The quantity inside the second sum is simply the probabilistic version of (4). For deterministic functions, information about the domain given a set of values in the range is given by inverse $f^{-1}$ applied to that set. In the probabilistic case, conditional probability plays the role of the inverse. 
We can move the first term inside the second sum, combine the summation as pairs and then it is not difficult to show $ \psi(\it P)$ simplifies to

$$\psi(\it P)= -H(X|Z)+{1 \over N}\sum_j H(X| r_j) $$

For the special case of identity transformation and for the partition $r_j={x_j}$, we can relate $\psi$ to  total correlation or multiinformation:

\begin{align}
\psi^I&= -H(X|X)+{1 \over n}\sum_j H(X| x_j) \\
             &={1 \over n}\sum_j H(X, x_j) - H(x_j)\\
             &={1 \over n}\sum_j H(X) - H(x_j)\\
             &= H(x_1, x_2,...,x_n)- {1 \over n}\sum_j H(x_j)
\end{align}

\noindent
On the other hand, total correlation or multiinformation of $X$ is given by

$$C=C(x_1,x_2,...,x_n)=\sum_j H(x_j)-H(x_1, x_2,..., x_n)$$

\noindent
Thus,

$$\psi^I =H(X)(1-1/n)-C/n  $$

\noindent
Here, we see that for the same total entropy, a large value of $\psi^I$ implies a small value for the
total correlation which implies higher degree of independence between variables.

Writing $X_D=X$ and $X_R=Z$ and rearranging terms to match  IIT notation, we have

$$\psi(\it P)= {1 \over N}\sum_k H(X_D| P_R^k)-H(X_D|X_R) $$

The need to identify the \lq\lq weakest" link, the decomposition into those parts that are most independent (least integrated)  has been raised in \cite {Tononi}. We can see that by construction $\psi (\it P)$ provides such a measure. For  bi-partitions a large value of $\psi (\it P)$ corresponds to a partition that breaks the \lq\lq weakest link". Of course, the problem of 
imbalanced partitions remains.  

$$\bar\Phi_{DM} [X;\it P]=\sum_{k=1}^N H(P_D^k|P_R^k)-H(X_D|X_R).$$

\noindent
IIT is essentially a constrained entropy clustering of pairs of random variables from the domain to range. The constraint is the fixed set of coordinates that prevents a Cartesian pairing of variables. A coordinate free measure should relax this requirement. We suggest a mechanism using 
$\psi$ here. 

Given the optimal partition, one which maximizes   $\psi (\it P)$, for a given partition class $N$ in the range, we can then split the domain by finding a matching $N$ partition by minimizing 

$$\sum_{k=1}^N H(P_D^k|P_R^k).$$

\noindent
Note the coordinates   are decoupled as $P_R^k$ was chosen first. Perhaps there exists a  metric or a normalization similar to IIT to define clusters or measures of complexity; the problem with imbalanced partitions remain in place. Direct
application of IIT with its normalization procedure defines a coordinate free measure.  
 We briefly give an example for decoupled partitions between range and domain described above; further numerical analysis is the subject of future research. Consider a cyclic permutation of independent variables $$ (x_1, x_2,....x_n) \rightarrow (x_n, x_1, ...x_{n-1})$$ If the partitions were not restricted coordinate-wise, the correct $n=N$ partition making each $H(P_D^k|P_R^k)$ zero is obvious. There is no interaction among variables and hence no integration. An easy check shows, IIT will miss the natural partition and produce $\Phi_{DM}>0$. One would expect fundamental measures to be exact on \lq\lq pure" sets of the theory {\it and} capture the underlying structures.

Overall, $\Phi_{DM}$ is  a useful measure of complexity for interacting systems in terms of simultaneous differentiation and integration.  There is no reason to believe consciousness requires only the highest degrees of simultaneous differentiation and integration by any particular measure; perhaps achieving consciousness only requires a critically high degree of complexity together with many other properties. Consciousness must be understood as a space-time transformation. Recent results \cite{Smith} show information patterns generated by the brain have fractal structure.  Statistical measures of cognitive complexity are likely to be defined at successive scales. Developing a coordinate independent multi-scale measure of integration and differentiation along the lines described above would be a possible direction.

\section {Casual Cognitive Networks}

{\it   Concepts of magnitude are only possible where a general
   concept is met with that admits of different determinate
   instances [Bestimmungsweisen].  According as, among these
   instances, a continuous transition from one to another takes
   place or not, they form a continuous or discrete manifold;
   the individual instances are called in the first case points,
   in the second case elements of the manifold.  -Riemann}\footnote{\cite{Riemann} translation by Rafael Sorkin.}

It is  fairly clear that neural events corresponding to consciousness are  distributed in space and time. The temporal spread of consciousness from  past into the future is a unique phenomenon and might be key in understanding its structure. 

Our point of view is that consciousness is a set of partially ordered {\it events}; 
The partial order is given by causal structure of neural activities. 
We model the partial order as a space-time directed graph of events where links transmit information 
from one computational event to another. The {\it longest} chain in this partial order  is a crucial object as we shall see later. 
The nodes which represent computational events combine inputs and produce outputs. Physics tells us information is never lost, but at classical scale of brain activity information does not include the dissipated heat or stray electrons; inference and computation are not reversible processes.  Is a null event, an event that produces no active output, part of structural complexity of cognition? We believe the answer is positive. A pattern of null events
has information implications because brain is an active system and a pattern of null and active events is detectable {\it in principle} whether it is actually tapped as output or not. 

Humans and animals who appear self-aware operate cognitively \lq\lq near present".  The spread in time raises several important questions. Could we, in principle, construct a cognitive system where the moments of perception or consciousness are spread over several minutes or several hours? Is there a limit to the temporal spread? Our usual intuition about a complex activity is that it requires substantial time  or at least more time does not diminish the results. The universe; however, shows us it does not always require more time in carrying out large scale transformations. This is for example suggested by time-energy uncertainty in Quantum Mechanics where the spread in time is limited by the spread in energy 

$$ \Delta T \Delta E \geq \hbar$$ 

\noindent
To achieve faster speed, one needs a larger spread in  energy which can be interpreted as a more complex information structure. Recent results show that entanglement, generally, in line with time-energy uncertainty, speeds up the evolution of Quantum processes \cite{Kupferman}. Inflation which must have involved the universe as a whole required very little time; in fact, the closer we get to the Big Bang where the universe must have operated as a highly integrated system,
we see a speedup in evolution of the universe. In the next subsection, we construct a set of plausible postulates. We do not claim it is the correct one. Different fundamental assumptions, for example based on continuity of consciousness in time,  are  possible and their consequences should be carefully considered.  The postulates are enumerated but we will discuss them without regard to order.  Central to these ideas
is the quantization of consciousness at  near present time intervals. 

\subsection { Postulates and their Consequences}

Suppose we manage to create a system composed of two brains intertwined in the same location 
of space where the information signals propagating throughout one brain is completely isolated 
(in the sense of information) from the other. Will one mind be aware of the other? or influence it 
in any way? Suppose we could scale up the brain so that the overall signaling and information 
processing would remain invariant but information signals  travel much longer 
paths. Will such a machine have the same quality of awareness as our own? Lastly, suppose we 
could freeze the activities of a machine which is in every aspect similar to the brain and start 
them again in time-steps years apart; processing the same information content, will consciousness bridge the time gap? 

Suppose a copy of the brain suddenly comes to existence at time zero. 
Suppose at the cross-section of time when consciousness is achieved first, the mind sends a 
signal to itself, directly or through memory paths, conveying the belief: \lq\lq what I think exists at 
present really belongs to the past". If consciousness were an accumulative stream then you would 
know (not through some abstract idea but through direct conscious perception) that what you see 
and perceive \lq\lq now" as an integrated whole, really belongs to a past no longer in existence. In principle, you would be 
able to delineate past from the future by a conscious mechanism while perceiving it as the 
\lq\lq present" a cognitive contradiction. Our mind, at best perceives the past as part of a transition. 
It is difficult even to imagine a moment of transition, the flow of water for example, lasting an hour.  
Consciousness is transient.  It is achieved by accumulation of computations then, it
rapidly collapses as the computational density drops and it switches from computation to communicating the results; We will attempt to produce a set of postulates below.

p1. Consciousness in quantized in time. 
p2. Consciousness is limited in the number of sequential computations. 
p3. Consciousness requires a minimum space-time computational density and must achieve a minimum total complexity.
p4. Consciousness is invariant under time contraction (in the sense of speeding up processes in the same reference frame) and transformation of communication which preserves time intervals and information content. 
p5. Consciousness is composed of decentralized integrating  networks  and peripheral asynchronous processes.
p6. Two peripheral processes associated with an integrating network are associated ( binding associativity) but integrating networks associated with overlapping processes are not necessarily themselves associated.
p7. Computational processes of  an integrating network are heterogeneous in time  with lower temporal boundaries associated  with information input and upper temporal boundaries associated with output.
p8. Events are binded (associated) by  {\it casual connections forward} in time and by {\it common cause backward} in time.
p9. Information flow across (p1) temporal boundaries is interpreted as a strict linear order and experienced as the arrow of consciousness.

It is not difficult to believe that if we \lq\lq run" the brain  (and its inputs) faster, we get the same result only in a shorter time period. Time dilation requires more thought. 

We propose that a cognitive process  cannot be broken into disjoint time segments.  We could approximate this segmentation by running our hypothetical cognitive machine for a short period, then slow it down to almost zero computational rate in the next time interval. Then run it at normal speed a second time. This has the effect of, essentially, turning the brain off and then on. Now imagine gradually speeding up the machine over the off interval. At some point, the three intervals will join. The sequential calculations achieve sufficient computational density to bridge the gap. This minimum density of sequential events is, according to our postulates, a universal constant (perhaps with some uncertainty laws or only in magnitude). It will be made precise in this paper. 
Note, we have indirectly limited the time interval over which consciousness can spread since the total number of allowed sequential events divided by the minimum required density gives us the maximum time.  We define idealized mechanisms by which more complex cognitive causal networks can be created from simpler structures with the same or fewer sequential steps.

One way of formalizing this idea constructively is the  following: Consider a network which generates certain class of computational events, the building block of a basic cognition machine. We can think of them as fundamental computational tasks some of which can be carried out concurrently and others sequenced. We can model these task relations via an undirected graph with a link between any two tasks which cannot be active at the same time because one task must generate a causal influence on others. We call this graph a {\it task influence graph}. It is clear that the minimum time to fire up events in these graphs is the chromatic number.  Now if we have another set of tasks with a different graph, we must look for a composition rule that does not increase the number of sequential calculations (colors). We know from graph theory that this can be done with tensor product of graphs. The composition mechanism of tasks into new  basic cognitive blocks must occur at time scales much faster than the total computation times of tasks themselves. It might require actual physical integration of modules carrying out these tasks or precise synchronization which
enables the composed modules to \lq\lq fire together".  We call the time scale at which these task compositions are carried out the   {\it Cognitive Planck Scale}. It is nearly an \lq\lq invisible" time scale as far as the computational structure is concerned.

The tasks can be roughly associated with  clusters of  neurons in a way that more active clusters across the brain 
influence less active neighboring  clusters (neighborhood is in the sense of communication distance); then another group from the set of previously  inactive clusters responds and so on. The clusters can be combined, by first replicating all of them a number of times, eventually requiring a brain with more neurons, and connecting them in a number of key places. The restriction is that the  combined pairwise clusters cannot take more time to produce a cognitive  event in their new combined configurations than they did before. 
 
 This process would eventually require the lowest building blocks to 
accept more connections to process larger number of inputs. It might also require faster communication channels across  clusters if
the clusters are not physically integrated but composed. 

Note that as far as computational density is concerned the computational action of  a  newly formed (composite)  cluster is seen as
a unified action, a single event. Nothing prevents the input information fields  from containing more complex information. The \lq\lq off-line" unconscious parts of the brain are free to use
as much sequential calculations and as much time as needed. It is only consciousness which must operate  near present.

Defining  the total number of  allowed computations to be multiple integers of the chromatic number would  permit us to describe the most basic oscillatory behavior of this idealized cognitive machine. In this context, our basic cognitive blocks defined above become {\it cognitive strings} and the
{\it reentrant }oscillation of these strings generate consciousness.  We can fit within the cognitive quantum several communicating strings of different frequencies, but the total time should not violate the maximal time spread.

Entanglement in QM requires that the composed system not be decomposable in terms of the Hilbert spaces of the sub-systems. This is done basically by addition; for photons with up and down polarization, {\it total state} \cite{Ulf} is given by:

$${1 \over \sqrt {2} } \left( \ket{+ }_1 \ket{- }_2 -\ket{- }_1 \ket{+ }_2 \right ) $$

Our first thought might be to find a way to \lq\lq add" graphs and this is probably an interesting graph theocratic problem. But,
our focus is on the chromatic number and we define {\it cognitive entanglement} as the collection of additions and subtractions of links of the tensor product of graphs that does not increase the chromatic number and prevents the decomposition of the graph back into
the product of its original components. Entanglement, here, is not a fine mathematical structure, just about any removal of links might lead to entanglement.

 We will return to this issue shortly after discussing some of the implications of the above simple model for the temporal structure of cognition. 

\subsection{Complexity of Cognitive Networks}

Consciousness is a transformation of information in time. A process which we have depicted by a network of causal events. 
It is {\it interpretive} in nature. In this paper, we are focused on architecture and space-time properties of cognition but make a few remarks about cognitive interpretation. We can model cognitive interpretation as particular types of concurrent computation. Finding the Kolmogorov instructions for  producing a restricted information set is interpretive (the focus is on the {\it process} of finding the instructions not the instructions Per se). We can view {\it classification} as an approximate
form of finding Kolmogorov instructions \lq\lq up to" certain classes. But, interpretation is not simply inference and removing redundancies. It also involves synthesis, innovation and imposition of patterns on information sets; it might involve for example  maximization based on certain class of symmetries by removing and adding additional information. Here is a good place to elaborate on peripheral processes. These are the \lq\lq upper" parts of our sensory systems and memory. They are
binded by the integration and interpretation network to the extent that they enter our the sphere of attention. Because cognitive  binding can be carried out backward in time (p8), cognitive integration has  (time limited) asynchronous properties. Note that we are focused on processes and not directly on physical networks; there is no exact or fixed association between neurons and these processes  analogous to the way tasks can be assigned in multi-core computers. This means, while one network is integrating sensory information, another can start a new {\it cognitive  quantum}. The key is that integration and interpretation  processes which are associated with network-wide production of outputs to memory and external organs, must be sufficiently apart in time that the output of an earlier quantum can provide integrable information for the later one.  So long as this rough linear order is maintained, we can define an {\it arrow of cognitive flow}, a linear order which is experienced as the unique flow of consciousness. 

In the above paragraph, we have hinted how the proposed cognitive  quantum is delineated in time. It begins with interpretation of an information set and ends with production of outputs (to external organs and memory). There is a gradual switch
from inputting information early in the process to outputting information later in the process; crucially, the end of a cognitive quantum is associated with rapid switch from computation to communicating  outputs. We can now connect this idea to computational density of p3. Earlier, we proposed that integration and interpretation processes fail if their density falls below a critical threshold.  It is indeed this same mechanism that delineates two cognitive quanta separating them by a \lq\lq soft" information boundary.

What happens if there is not enough information flow from one cognitive quantum into another, for example when the rapid drop in computational density happens simultaneously in two integration networks? Then two independent arrows emerge corresponding to two different minds. We know this happens because different people have different cognitive arrows. There are indications that the split can occur within one brain as well \cite{Dominus}-\cite{Harrington}. Such splits might happen more often than introspection leads us to believe. Two such event clusters might
overlap in space but once computational density drops, a temporal boundary is formed which
can transmit information to the future events but, it cannot be used to access the events already experienced; the past has finally become inaccessible. Cognition just like the physical world has a notion of inaccessible past. While the cognitive quanta might give awareness a \lq\lq discrete" \cite{Koch2} temporal structure, the existence of concurrent quanta enable the brain to process any incoming input.  

We can now explain how we perceive the outer world using our model. A good example is the video and voice synchronization. people are capable of compensating up to about 100 milliseconds of delay between audio and video.  If the delay is small enough both types of sensory information can be placed inside the same cognitive quantum where physical  time does not exist and our mind can create any temporal information pattern which is useful. When the delays is large, brain has to place some voice and sound data into one quantum and others into the next. One quantum can begin before the complete end of the previous one. The discordant voice and video data in the same quantum leads to competition \cite {Baars1}.

 A little thought shows that we have arrived at a distributed version of Global Workspace Theory \cite{Baars2}. The mind is
a theater with some (potentially conscious) events occurring on the stage and others on the backstage. There are multiple \lq\lq spotlights" in various degrees of  \lq\lq turning on". One of them  reaches a critical threshold turns brightly and outputs a scene on the stage. It then turns off.  Another spotlight turns on a short while later (with varying but bounded temporal delay). The audience somehow
never perceives the darkness. Typically, but not always the spotlights shinning at two different places do not turn on at the same time, giving one brightened scene enough time to substantially influence what the audience sees next.

We have elaborated on the meaning and implications of all postulates except parts of p4. There are two types of invariant transformations when we consider the embedding of cognitive events in spacetime. One comes directly from symmetries of physical  space. For example translation in space leaves cognition invariant. Another type is what we call {\it conformal cognitive transformations}. Think of two computers connected by a cable. We know it does not matter if the cable is straight, twisted or has any other shape. Now suppose we disconnect parts of Corpus Callosum and connect the ends to a hypothetical device which carries cognitive information, at speed of light, kilometers away then turns around and carries the information from one hemisphere into another, the content of the data and time intervals remaining exactly the same. p4 implies that consciousness remains invariant. The class of space-time transformations which leave consciousness intact merits  investigation. In General Relativity, geodesics are followed as determined by the equation: 

$$ G_{\mu\nu}\Lambda g_{\mu\nu} = {{8 \pi G} \over {c^4}} T_{\mu\nu}$$

\noindent
An  equality between local geometry of spacetime and local dynamical information; the rules of the evolution of the system are entirely determined by  this relation. As mentioned earlier, the
causal structure of events contains much of the geometric information in curved spacetime.
Subjective space is closely related to concurrency of binded events. For example
our subjective visual perception can be mapped onto the Primary Visual Cortex (V1) where a damaged section would lead to the perception of a missing area in the visual field. 
A conformal cognitive transformation, one which produces exactly the same computational events with the same temporal relations would produce the same subjective visual field. We could move
some of the neurons in V1, permute their positions but so longs as their causal event relation with each other and with other neurons does not change, we still see the same visual field.
 The cognitive coupling of space and time  is manifest in our subjective experience of motion. 
Here too, it is causality that determines subjective geometry. The actual geometry of physical space does not play a direct role; it is timing and causality that matter. The idea that physical spacetime emerges entirely from dynamical information of events has been expressed in \cite{Smolin1} for example. We will make more comments about this subject at the end of this paper. 
But what are fundamental cognitive events? We will answer this question shortly as well.
 
The above arguments are independent of the exact role of V1 in conscious awareness which remains the subject of current research
\cite{Tong}-\cite{Leopold}. The main point is that subjective perception of space and time  depend on causal properties of events embedded in physical reality.  Although we do not yet know with any certainty at what point conscious processing of information begins in the hierarchy of brain activities, for example whether events in V1 are 
involved or not in visual awareness, there must be a \lq\lq boundary" between neural regions responsible for unconscious preparation of information and the
\lq\lq neural correlates" of consciousness. This 
boundary is fluid and can dynamically extend to some degree into \lq\lq upper" parts of sensory and memory circuits depending on their interactions with the integration and interpretation networks.
The integration and interpretation networks are  subject of very active current research  and might involve 
thalamus, frontoparietal circuits,  prefrontal cortex, amygdala and even cerebellum (Many studies have also demonstrated crossed connections between the dentate nucleus and the dorsolateral prefrontal cortex in the cerebellum-thalamic-cortical direction) 
\cite{De}.  Later in this paper,
we will remark on hierarchies of feedback and control which might be important to consciousness. Higher level neural activities  
which in part correlate  with consciousness do not enter an inactive mode when brain is in \lq\lq default" mode, a state in which we  show little focus on sensory inputs. The default network activity may be driven from highly coupled areas of the posterior medial and parietal cortex, which in turn link to other highly connected and central regions, such as the medial orbitofrontal cortex \cite{Hagmann}. Remarkably, the brain requires little extra energy \cite{Raichle} for engaging in a particular task, only about \% 5 percent. 
Brain's integrated network of causal events always has abstract \lq\lq equivalent"  sensory information within it with little information content. We can consciously  \lq\lq see" total darkness and hear a constant monotone sound; high levels of sensory content does not seem to be a requirement in these cases. The neural activity patterns maintain high degree of complexity regardless. Think of differentiation acting on a constant function which does not make differentiation  itself any less complex of an operation. When sensory information becomes available, these activities are reconfigured back to their original purpose and expanded as consciousness extends into the upper levels of sensory networks themselves. There can never be a quantum of consciousness without integrated  complex modules.  Each module represents highly connected collection of sub-transformations (Consider a fast Fourier transform where each coefficient is obtained from the same data set and uses results from steps for calculation of other coefficients) that produce output in a synchronized fashion. Indeed, it is this {\it nearly} simultaneous completion of a set of laterally connected transformations (each composed of highly integrated sub-transformations) which constitutes a unified moment of consciousness.

We return to the simple graphical representation of cognitive events using tensor products next.

Since there is evidence that neural activity patterns and neural connectivity structures are fractal, we looked for a simple computational model that generates self-similar structures according to compositional rules which increase complexity but not overall number of computational steps and that can represent a pattern of activation and inhibitory   connectivity. 

Perhaps, because sensory processing such as smell, touch and vision required concurrent processing and because  early organisms had to operate near present, concurrent processing became an essential part of the development of the nervous system. Therefore, consciousness relied on this  concurrent architecture as it developed.

The tensor product \cite{Moradi} of two connected graphs  where both graphs are not bipartite is connected; these are reasonable assumptions for modeling complex cognitive networks.

Computation and communication can be represented by directed acyclic graphs generated by such networks. Cognitive systems are complex; what kind of complexity and to what degree remains unclear. But, we should expect cognitive composition of networks to increase complexity.  Many of graph complexity measures require that the number of the edges be given by particular powers of the number of vertices, for example, the geometric average between the number of edges of simple chain graphs and complete graphs: $n \sqrt{n} $ ( $n$ is number of vertices). This relation is not  preserved under tensor product operation where the number of edges is twice the product of number of edges of two graphs while the number of vetices is the product of number of vertices. 

If we have two simple graph $G_1(n,l )$ with $n$ vertices and $l$ edges, and another $G_2(m, k)$. The number of edges of the tensor product graph is $2kl$:

$$ G_{1 \otimes 2}(mn, 2kl)=G_1(n, l)\otimes G_2(m, k). $$

But, $ nm \sqrt{nm} = n \sqrt{n}\cdot m \sqrt{m}$ while the number of edges would be  $ 2 n \sqrt{n}\cdot m \sqrt{m}$ (for $k= n \sqrt{n}$ and $ l= m \sqrt{m}$).

However, entanglement as defined above involves removal of a net number of  edges, say $r$ which cannot increase the chromatic number $\xi$ \cite{Duffus}:

$$\xi (G_{1 \otimes 2}(mn, 2kl)) \le min \{\xi( G_1(n, l)), \xi( G_2(m, k)) \}$$

$$\xi (G_{1 \otimes 2}^{*}(mn, 2kl-r)) \le \xi (G_{1 \otimes 2}(mn, 2kl)) $$

\noindent
Note that if we add and delete the same number of edges the total number is unchanged. So the pruning mechanism might involve
such neutral adding and subtracting edges. Repeated tensor product composition of basic blocks followed by pruning leads to an approximate self-similar structure which is consistent with space and time fractal properties of neural measurements. We are not suggesting cognitive causal networks are exactly
the most complex networks by some fixed complexity measure but that they have high degree of complexity and that this complexity is 
maintained by tensor products without increasing the chromatic number.

\subsection { Cognitive Binding }

According to our postulates both common cause and common effect bind events. This means, in graph language, 
events can be binded by graph connectivity measures of the space-time events regardless of direction of edges.
There might be a cost in going backward and forward in time when connecting events. It might be that only
paths with few oscillations in time can contribute to the binding mechanism (similar to Feynman's path integral formulation of QM in a different context). At this point we do not have a formulation for this idea.  

 The existence of causal influence is necessary but not sufficient for binding events. Causal chains are the \lq\lq skeleton"
 of a cognitive event network but more is required.  
The events themselves must form coherent subsets. We do not consider this problem fully in this paper.
 But, the integration and interpretation networks have one obvious geometry. Early in the process they accept inputs and toward the end produce more outputs. The network has to have the right global 
properties. Another geometric property  might be dynamic feedback. Geometrically these are edges which bypass several computational steps crossing over time and reaching future events. Connecting two graphs along their \lq\lq boundaries"  is probably less effective in generating complexity (if we think of complexity in terms of non-isomorphic subgraphs) than connecting them at multiple points. 
The range of direct feedback is limited as the past becomes irrelevant behind a barrier of cognitive  innovation and inference. We could say that cognitive processes have limited Markov order; they are not  small world in the time direction. This too would be a geometric property: direct forward edges cannot be \lq\lq long". 

Hierarchy  appears to be a fundamental geometric property of cognitive systems.  The brain is capable of \lq\lq looking inward" and does not require  sensory information or even specific memories to remain conscious. The integration of sensory information followed by possible actions is seamless because the brain is always ready to accept and process them \cite{Raichle}. 
As neural layers  became more complex thorough biological evolution, the neural patterns utilized to process and control information from sensory networks, themselves could have become subject to information processing. By analogy consider a network of agents processing information, such as a group of people trying to piece together a big puzzle. At some point it might be efficient to treat the activities of the people themselves as pieces of a puzzle at higher level in order to apply  supervisory policy control over their activity patterns, a holistic feedback control mechanism which treats the ensemble of controllers as a higher information \lq\lq field". If the state-space of controllers or agents can be given a manifold structure, for example by methods of {\it information geometry} \cite{Amari}, parametrization of paths on the manifold can be viewed as temporal patterns corresponding to a particular sequence of evolving policies. In other words, the binding problem is not simply a matter of finding  multi-scale connectivity metrics defined locally but producing the right {\it cognitive motifs} and associating them in specific ways to    each other in a network of space-time events. These ideas are subject of future research.

\subsection {The Fundamental Cognitive Chain}

Our idealization with tensor product composition, by construction, limits the spread time. It is then straightforward to
define failure under time dilation because of reduction in density. 

If we allow unbounded chain of computational events, there cannot be a limit to temporal spread of consciousness. This is of course
a possibility but in this paper and our postulate system, we are exploring intrinsic mechanisms which would limit all self-aware
cognitive systems to operate in  near present regardless of their complexity. 

p2 asserts the number of sequential computation is limited. To understand this limit, we will first construct the fastest possible
cognitive system equivalent to the simplest cognitive system which first achieved consciousness, perhaps a vertebrate. Consider a conscious \lq\lq moment" of such a system; our machine must emulate this moment. 
We could begin by bringing the neurons closer, perhaps the axons could be rearranged to take shorter paths. Much of the
life support mechanisms  of the cells is not involved in cognition. We replace the cells with equivalent computation machines without such overheads. This device will be macroscopic. There exists a limit to temporal miniaturization. In this device every computation is essential, communication is done the fastest possible way laws of physics allow, perhaps with photons at speed of light and we assume it produces little overhead and crucially, events which correspond to mere transmission of information are detectable. We now focus on the cognitive portion of events in this brain which corresponds to consciousness. We consider the {\it longest} possible chain of multi-body fundamental
 interaction\footnote{The count number of
such interactions, even counting the number of particles in a region of space is subject to uncertainty laws of physics. Since we cannot discount Quantum Field effects at very small scales, to truly count the \lq\lq events", we must assume a discrete event structure underlying the universe along the lines of Causal Sets theories. It is interesting that in Causal Sets theories, the longest chain of events is a geodesic and related to {\it proper time} in General Relativity.   Even within classical electromagnetism, we can have ensemble of photons flying through vacuum solve  Fourier transforms. How many events are in such classical  continuous flow? If the device uses  Bose-Einstein condensate, how do we count multi-body interactions? Alternatively, we could assert that all cognitive systems are equivalent to classical Universal Turing Machines, then we could construct the equivalent minimal  machine and count the longest sequential chain.  Quantum computation cannot decide the undecidable classical computational problems.} This is a fundamental number $c_{max}$. It cannot be increased. What is the difference between a human and say a self-aware primate? The human minimal device has more concurrent events, each event involves more particles, it has more particles flying around, more photons and electrons but the
longest chain of computational events is the same. 

Additional interactions solely to maintain and transmit state information constitute a reduction in density, we assume fundamental laws recognize them as such. We can now slow down these miniaturized ultra-fast brains. As we slow them down, the computational density $d=c_{max}/time$ decreases and the length of the longest causal path increases. There is a critical threshold $d_{min}$ below which consciousness disappears. 

Alternatively, we could use the  \lq\lq volume" of the weakest cognitive machine producing consciousness to limit the time spread. This volume will fail under time dilation at some critical density. All other cognitive machines fail in producing consciousness at about the same time spread.

Our postulates imply  not just the original consciousness but no other consciousness would exist 
after time dilation. Can we justify the postulate? One strong justification might be the general idea that the universe demands efficiency and there is a limit on how inefficient a process can be in producing an emergent fundamental entity. Causal distance
growing with the number of interactions, lengthening the binding distance also points to a limit. 

A more subtle approach might be to weaken the postulate and say, only, that the original consciousness disappears.
This leaves open the possibility that a new  consciousness is produced. It can come about the following way:
Consider the very last few milliseconds of the time dilated process. What we have is some state of cognition which
must undergo effective transformation in those last few milliseconds; otherwise, the process would have been complete earlier.
Then, from that state forward in time, perhaps we can have a new cognitive quantum. The state embodies the computations as if they were done instantly. Then, time spread would be measured from
that state forward for just a few milliseconds. The original intent of the postulate has not been violated; computational density would 
in fact be even higher and efficiency holds. But, there is a subtle obstacle here. Imagine cognition as the process of turning a photograph into a cartoon, an interpretation. This is not far from how we actually recognize faces by feature extraction. The features though are not all extracted
at one time. What we think we see within a moment of consciousness are different key features extracted over time. 
Think about the Kolmogorov instructions for reducing an object, like a circle. We can calculate its center from three points, then we measure the distance from the center to one of the points. We do not need to wait for the calculation of the radius to write into memory the coordinates for the center which was obtained first. Consciousness  is the transformation of information itself which includes the outputs (whether they are written into memory or not, the point is that they could be in principle). We have assumed in our postulates that the latter part of a consciousness moment  is output oriented, similar to the drawing of a cartoon speeding up as some general basic features are produced, but every {\it non-invertible} process is potentially a feature and the transformation is what lies  between the first non-invertible integrating computation and its completion\footnote{ This point of view supports the idea that most of V1 is not involved in visual awareness; there is little information loss as a result of integration early in the process and hence there is little  \lq\lq visual interpretation". } The high level, more abstract processes which produce our self-awareness operate the same way. Consciousness is not a state or an output; it is a {\it behavior}. It is the  internal observation, control, inhibitions and competitions  that produce those outputs. They are not embodied in the terminal few milliseconds seconds. Imagine the state being
maintained for a century and moved to another location in the solar system. Would its cognitive transformation in the last few milliseconds suddenly rush back in time find the entire process and behavior which gave rise to it and become conscious? More likely, it would be just a given state at that time transformed to produce some output, including for example perhaps parts of an image, devoid of a subjective experience. We give a further philosophical rational in terms of emergent realties and time scales in at end of this paper.

\subsection{ Unbounded Temporal Spread} 

Although not the focus of the present paper, we briefly consider alternative ideas in this section.  Evolutionary considerations point to consciousness as having a purpose and that purpose is probably 
production of useful integrated  output from highly pre-processed disparate initial data.  However,  our cognitive experience is  poorer than
it appears. Our highly detailed precise visual field is quite small. Storage of information in visual short-term memory (VSTM) a key component of many complex cognitive abilities is highly limited in capacity \cite{Lara}( VSTM is involved in holding a visual object in conscious awareness \cite{Pun}). Sequential information whether auditory 
information or linguistic  quickly passes through our mind. 
Yet, disparate  information processes appear to easily enter our subjective experience concurrently, all of our five senses, memories, emotions and sense of self can be experienced within the same \lq\lq moment" of consciousness. 
Multi-scale interpretation of information would require not just concurrent computation but significant sequential processing in order to make sense of relations and correlations at various scales. The idea of fundamental chain puts  a uniform limit on the ability of
cognition to essentially act as a \lq\lq time machine" bringing events of the past vividly into present. It also
explains why our awareness  of the outer world is expansive yet lacks richness, relativity speaking, as Dan Dennett has pointed out.

If we do not place a limit on spread time or number of sequential computations, it seems that in principle, the spread time can be unbounded, enabling a cognitive system to integrate large information sets in space and time. What could, in principle, prevent the existence of a self-aware cognitive system, not necessarily a biological one, 
from having a \lq\lq retina" several kilometers in diameter? (Obviously a large crowd of people
do have a large total visual surface which can cover a large field but they cannot see the large scale patterns.)  Processing an information rich visual field would require more sequential calculations. Complexity theory in Computer Science makes it clear we can create information patterns whose processing time grows with the size of the problem, requiring more sequential calculations; there is a limit to parallel computation. What part of information processing of the brain cannot be scaled up in space and time Per se? We might resort to practical obstacles. In analogy, stacking up wooden cubes, one on top of each other does not have 
a mathematical limit but in practice quickly becomes vulnerable to perturbation. This  indeed might be the case but, information processing is abstract and 
can involve error correction, repetition and  access to memory. If  there are no fundamental laws of the type developed in this paper to limit temporal spread,
we should consider the possibility of cognitive systems which can unify past and future events for much longer times than milliseconds. Even in that case, we believe a lower limit on computational density remains. The density will not be computed across a single causal chain  but across a temporal cross section, allowing concurrent computation in space to be aggregated to increase the density. In this case, the more powerful the brain, the more it can slow down through time with consciousness intact.

\subsection{ Field Theories of Consciousness and  Physical Laws}

An interesting paper on the relation between consciousnesses and the inflation process of early universe gives a good summary of
QM theory of the mind \cite{Zizzi}:

\lq\lq In a brain's neuron there is the cytoskeleton, which is made of protein networks. The most important components of the cytoskeleton are microtubules. Microtubules are 
hollow cylindrical polymers of proteins called tubulins. Tubulins are dipoles and they 
can be in (at least) two different states (or conformations). Tubulins have been studied in classical computing. In fact simulations suggest that tubulins behave as a classical CA. But tubulins can also be in a superposition of the two (or more) conformation states. 
In this case they are qubits, and they behave as a biological quantum cellular automata. 
Indeed, tubulins can perform both classical and quantum computing."

The arguments and calculations in \cite{Tegmark} are  convincing that the computational events of the brain
are classical. But the field theories, QM  or electro-magnetic  might play an important role in the binding problem that should be considered.  Perhaps, the coherent  fields generated by large regions of the brain appear as random but structured global \lq\lq noise" to processes of casual events and might be part of a multi-layer binding mechanism, an additional stochastic connectivity pattern. If so, dampening field induced noise in AI systems might be detrimental to emergent of consciousness. This also implies the feedback loops earlier mentioned are intrinsically stochastic control problems on information manifolds.

The relation between consciousness and fundamental laws of physics  could be examined in the opposite direction as well. Our mind  constructs subjective notions of space and
time from a partial causal order. These notions reflect the classical properties of spacetime but are emergent. We have suggested that space is not directly crucial to consciousness, the important agent is time as it relates directly to the causal structure of events. We gave the example of the communication between the two hemispheres reconfigured such that  the modified brain can have radius of hundreds of kilometers with cognition processes remaining unchanged. However, human brain has a particular overall shape and is highly organized in three dimensions. These basic observations might have implications for background independence property of physical laws.   Causal Set Theory \cite{Sorkin} is based on the minimal idea of partial order without assuming any topology or geometry. Loop quantum gravity can be considered as fixing a particular topology \cite{Asht}. String Theories lack a full non-perturbative definition in arbitrary space-time background. Consciousness  is a level of reality with its own fundamental properties and laws embedded inside but distinct from physical reality. Its subjective concepts of space and flow of time are, nevertheless, based on the {\it proto-geomtry} (from the perspective of consciousness) of physical spacetime. Our universe might be embedded inside a much larger state-space, a more primitive but expansive reality. Thus, the nature of the underlying background might be at least two layers deep and various fundamental theories might be targeting different depths of it. The main challenge  may not be the construction of a background independent theory but understating how the background properties of one reality are embedded  in a deeper one. 

It has been suggested that perhaps complex systems such as echo systems or Internet\footnote{The Nature of Consciousness: How the Internet Could Learn to Feel, The Atlantic, interview with C. Koch} are consciouses or can achieve  consciousness if they become sufficiently integrated. Consciousness has emerged out of the physical reality at classical scales; temporal causality and complex composition are essential elements of emergence. But, if a higher reality is to emerge from and above consciousness, it might be required to utilize the lower reality as building blocks, the same way our brain uses particles and fields of physical reality. The proto-type of such a higher reality is history  existing on much longer time scale.  Motifs of this reality are  emergent in human interdependent phenomena ranging from  mythology to culture to  global economy.
History too has a fabric of human interactions which are most integrated and exercises emergent executive control over human populations that provide it with its forward energy. History also can be understood in quasi periodic epochs where most rapid and most influential changes take place and are recorded into collective long term memories.  
History is sensitive to the actions of a single individual while some others are peripheral to it. But, this higher level of reality, even if it exists, might not be accessible to to our consciousness, the same way our consciousness  while embedded in physical reality is sharply separated from it.  These ideas about history  were first most clearly expressed by Fredrick Hegel. Hegel believed opposing elements and ideas that interact do not annihilate one another but together forge a new  configuration   that can only be understood at the end state and as the entire history of that composition.\footnote{ \it \lq\lq die Eule der Minerva beginnt erst mit der einbrechenden Dammerung ihren Flug"}  It might be argued that human consciousness already represents such a transition relative to proto-consciousness present in more primitive brains of animals, allowing its cognitive quanta to operate at much larger time scales; however, human mind appears to operate near present  not unlike  brains of other vertebrates. 

\section*{ Acknowledgments} 

The author is grateful for helpful comments from  Gabriel Lipsa, Scott Watson, Bernard J. Baars and Rafael Sorkin.

\end{document}